\documentclass[11pt]{article}
\usepackage{amsfonts}

\newcommand{\set}[1]{\left\{ #1\right\}}
\newcommand{\gilt}{:}
\newcommand{\sodass}{\,:\,}
\newcommand{\setGilt}[2]{\left\{ #1\sodass #2\right\}}

\newcommand{\realrange}[2]{\left[#1, #2\right]}

\newcommand{\unitrange}[2]{\realrange{0}{1}}

\newcommand{\llabel}[1]{\label{\labelprefix:#1}}
\newcommand{\labelprefix}{} 

\newcommand{\discussionsize}{\small}

\newcommand{\frage}[1]{}

\newenvironment{code}{\noindent
\begin{tabbing}%
\hspace{2em}\=\hspace{2em}\=\hspace{2em}\=\hspace{2em}\=\hspace{2em}\=%
\hspace{2em}\=\hspace{2em}\=\hspace{2em}\=\hspace{2em}\=\hspace{2em}\=%
\kill}{\end{tabbing}}

\newcommand{\labelcommand}{}
\newcommand{\captiontext}{}
\newsavebox{\codeparam}
\newcounter{lineNumber}
\newenvironment{disscodepos}[3]{%
\renewcommand{\labelcommand}{#2}%
\renewcommand{\captiontext}{#3}%
\sbox{\codeparam}{\parbox{\textwidth}{#3}}%
\begin{figure}[#1]\begin{center}\begin{code}\setcounter{lineNumber}{1}}{%
\end{code}\end{center}\caption{\llabel{\labelcommand}\captiontext}\end{figure}}

{\end{disscodepos}}

\newcommand{\Is}       {:=}

\newdimen\endofsize\endofsize=0.5em
\def\endofbeweis{~\quad\hglue\hsize minus\hsize
                 \hbox{\vrule height \endofsize width
\endofsize}\par}

\usepackage[utf8]{inputenc}
\usepackage{algorithmic}
\usepackage{tabularx}
\usepackage{algorithm}
\usepackage{amsmath}
\usepackage{amssymb}
\usepackage{color}
\usepackage{fullpage}     
\usepackage{latexsym}
\usepackage{makeidx}
\usepackage{multicol}
\usepackage{numprint}
\usepackage{t1enc}
\usepackage{times}
\usepackage{graphicx}
\usepackage{url}
\usepackage[left=2cm,top=1.5cm,right=2cm]{geometry}
\npdecimalsign{.} 

\definecolor{mygrey}{gray}{0.75}
\newcommand{\ie}{i.e.\ }
\newcommand{\etal}{et~al.\ }
\newcommand{\eg}{e.g.\ }

\def\MdR{\ensuremath{\mathbb{R}}}

\newcommand{\mytitle}{ {\color{red}KaHIP} v3.00 -- {\color{red}Ka}rlsruhe {\color{red}Hi}gh Quality {\color{red}P}artitioning \\ User Guide}
\begin{document}
\title{\mytitle}
\author{Peter Sanders and Christian Schulz\\ 
	\textit{Karlsruhe Institute of Technology}, \textit{Karlsruhe, Germany} \\
	\textit{University Heidelberg}, \textit{Heidelberg, Germany} \\
	\textit{Email: \url{sanders@kit.edu}, \url{christian.schulz@informatik.uni-heidelberg.de}} }
\date{}

\maketitle
\begin{abstract}
This paper severs as a user guide to the graph partitioning framework KaHIP (Karlsruhe High Quality Partitioning). We give a rough overview of the techniques used within the framework and describe the user interface as well as the file formats used. Moreover, we provide a short description of the current library functions provided within the framework. Since version 3.00 we support multilevel partitioning, memetic algorithms, distributed and shared-memory parallel algorithms, node separator and ordering algorithms, edge partitioning algorithms as well as ILP solvers.
\end{abstract}
\paragraph*{Project Contributors:} Yaroslav Akhremtsev, Roland Glantz, Alexandra Henzinger, Dennis Luxen, Henning Meyerhenke, Alexander Noe, Wolfgang Ost, Ilya Safro, Peter Sanders, Sebastian Schlag, Christian Schulz,  Daniel Seemaier, Darren Strash, Jesper Larsson Träff
\tableofcontents
\thispagestyle{empty}

\vfill
\pagebreak
\section{Introduction}
\begin{figure}[h!]
\begin{center}
\includegraphics[width=8cm]{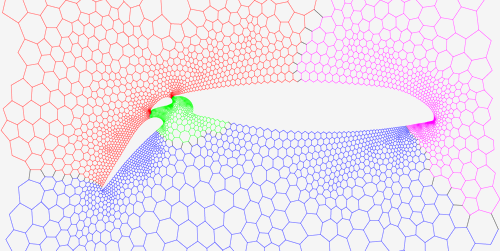}
\end{center}
\caption{An example mesh that is partitioned into four blocks indicated by the colors.}
\label{fig:airfoil}
\end{figure}

Given a graph $G=(V,E)$ and a number $k>1$, the graph partitioning asks for a division of the graphs vertex set into $k$ disjoint blocks of roughly equal size such that some objective function is minimized. The most common formulation minimizes the number of edges that run between the blocks. An example is given in Figure~\ref{fig:airfoil}. 
Nowadays, the graph partitioning problem has many applications in different areas such as parallel scientific computing or graph computations \cite{schloegel2000gph,catalyuerek1996dis,zhou2010controlling,heuvelinecoop,bomansc13,bulucc2012graph,SalihogluW13}, route planning \cite{Lau04,wagner2005pgs,SWZ02,ls-csarr-12,klsv-dtdch-10,MSM07,DellingGPW11,DellingW13}, VLSI Design \cite{alpert1995rdn,alpert1999spectral,vlsi-ph-design,vlsicad-book} or in solving sparse linear equation systems \cite{george1973nested}\footnote{Let us know if we have  missed your application here!}. There has been a vast amount of research on graph partitioning. We refer the reader to \cite{SPPGPOverviewPaper} for more material on graph partitioning. 

{\color{red}KaHIP} - Karlsruhe High Quality Partitioning - is a family of graph partitioning programs based on the publications \cite{kaffpa, kaffpaE, kabapeE,kappa,SafroSS12,highqualitygraphpartitioning,dissSchulz}. It includes KaFFPa (Karlsruhe Fast Flow Partitioner) in its variants Strong, Eco and Fast, KaFFPaE (KaFFPaEvolutionary) which is a parallel evolutionary algorithm that uses KaFFPa to provide combine and mutation operations, as well as KaBaPE which extends the evolutionary algorithm. 
Moreover, we include algorithms to output a vertex separator from a given partition.

KaHIP focuses on a version of the problem that constrains the maximum block size to $(1+\epsilon)$ times the average block size and tries to minimize the total cut size, \ie the number of edges that run between blocks. 
To be more precise, consider an undirected graph $G=(V=\{0,\ldots, n-1\},E,c,\omega)$ 
with edge weights $\omega: E \to \MdR_{>0}$, node weights $c: V \to \MdR_{\geq 0}$, $n = |V|$, and $m = |E|$.  We extend $c$ and $\omega$ to sets, \ie 
$c(V')\Is \sum_{v\in V'}c(v)$ and $\omega(E')\Is \sum_{e\in E'}\omega(e)$.
We are looking for \emph{blocks} of nodes $V_1$,\ldots,$V_k$ 
that partition $V$, i.e., $V_1\cup\cdots\cup V_k=V$ and $V_i\cap V_j=\emptyset$
for $i\neq j$. The \emph{balancing constraint} demands that 
$\forall i\in \{1..k\}\gilt |V_i| \leq (1+\epsilon)\lceil\frac{c(V)}{k}\rceil$ for some imbalance parameter $\epsilon$.
The objective is to minimize the total \emph{cut} $\sum_{i<j}w(E_{ij})$ where 
$E_{ij}\Is\setGilt{\set{u,v}\in E}{u\in V_i,v\in V_j}$. 
Methods to optimize the maximum communication volume of a partition are also included in the package. There is also a version of the program that enables to balance nodes and edges between the blocks which is in particular important for applications in which workload is associated with the edges as well such as sparse matrix vector multiplications or in graph processing frameworks. In this case, we set the node weight to $c(v)+\text{deg}_\omega(v)$ for a node $v\in V$ and set the balance constraint to $\forall i\in \{1..k\}\gilt |V_i| \leq (1+\epsilon)\lceil\frac{c(V)+\sum_v \text{deg}_\omega(v)}{k}\rceil$.

The purpose of the manual is to give a very rough overview over the techniques used in the partitioning programs, as well as to serve as a guide and manual on how to use our algorithms. 
We start with a short overview of the algorithms implemented within our graph partitioning framework. 
This is followed by a description of the graph format that is used. 
It is basically the same graph format that is used by Metis~\cite{karypis1998fast} and Chaco~\cite{chaco}, and that has been used during the 10th DIMACS Implementation Challenge on Graph Clustering and Graph Partitioning~\cite{dimacschallengegraphpartandcluster}. We then give an overview over the user interface of KaFFPa, KaFFPaE and KaBaPE and explain which program to use to derive node separators. We finish this manual with the description of the library functions provided by the  current release.
\vfill
\pagebreak

\section{Graph Partitioning Algorithms within KaHIP}
\begin{figure}[b!]
\begin{center}
\includegraphics[width=.75\textwidth]{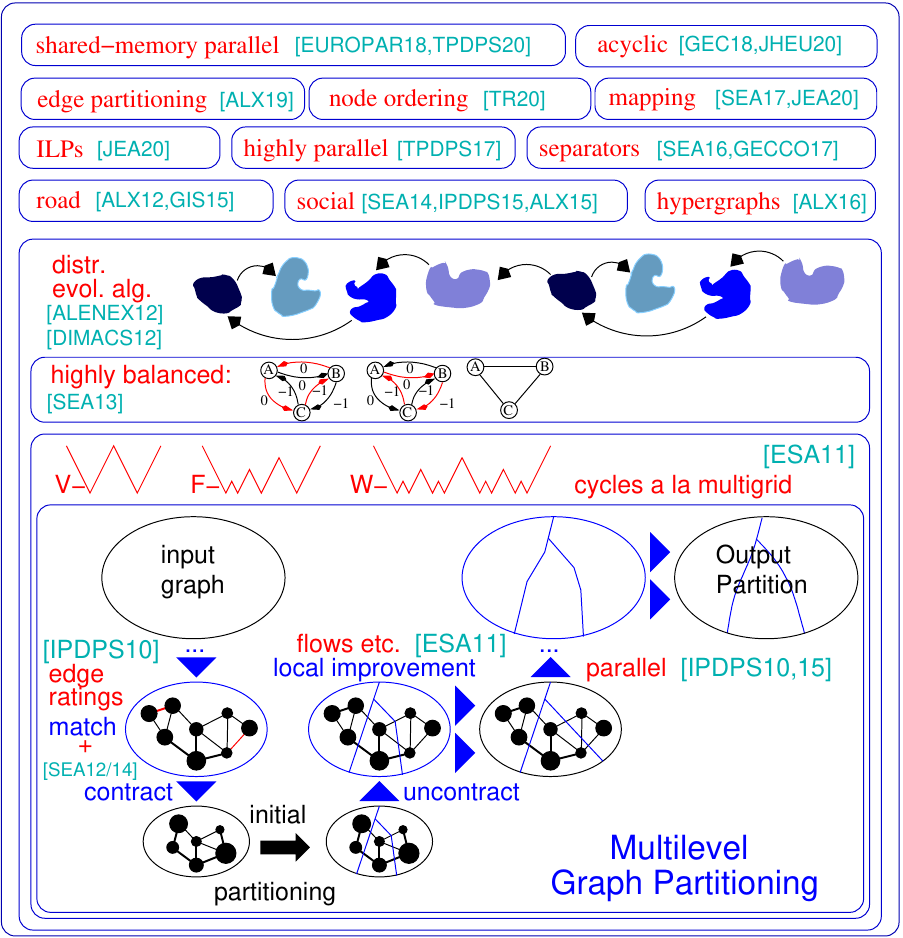}
\end{center}
\caption{Overview over the techniques used in the KaHIP graph partitioning framework.}
\label{fig:MGPallen}
\end{figure}

We now give a rough overview over the algorithms implemented in our graph partitioning framework. For details on the algorithms, we refer the interested reader to the corresponding papers. Figure~\ref{fig:MGPallen} gives an overview over the techniques used with the framework.\\
\subsection{KaFFPa}
A successful heuristic for partitioning large graphs is the \emph{multilevel graph partitioning} approach, where the graph is recursively \emph{contracted} to create smaller graphs which should reflect the same basic structure as the input graph. 
After applying an \emph{initial partitioning} algorithm to the smallest graph, the contraction is undone and, at each level, a
\emph{local search} method is used to improve the partitioning induced by the coarser level. 
KaFFPa is a multilevel graph partitioning framework which contributes a number of improvements to the multilevel scheme which lead to enhanced partitioning quality. 
This includes flow-based methods, improved local search and repeated runs similar to the approaches used in multigrid solvers.

Local improvement algorithms are usually variants of the FM algorithm \cite{fiduccia1982lth}.
The variant KaFFPa uses is organized in rounds. In each round, a priority queue $P$ is initialized with all vertices that are incident to more than one block, in a random order.
The priority is based on the gain $g(i) = \max_P g_P(i)$ where $g_P(i)$ is the decrease in edge cut when moving $i$ to block $P$.
Local search then repeatedly looks for the highest gain node $v$ and moves it to the corresponding block that maximizes the gain.
However, in each round a node is moved at most once.
After a node is moved, its unmoved neighbors become eligible, i.e. its unmoved neighbors are inserted into the priority queue.
When a stopping criterion is reached, all movements after the best-found cut that occurred within the balance constraint are undone.
This process is repeated several times until no improvement is found.

\paragraph*{Max-Flow Min-Cut Local Improvement.}  KaFFPa additionally uses more advanced local search algorithms.
The first method is based on  max-flow min-cut computations between pairs of blocks, in other words, a method to improve a given bipartition.
Roughly speaking, this improvement method is applied between all pairs of blocks that share a non-empty boundary.
The algorithm constructs a flow problem by growing an area around the given boundary vertices of a pair of blocks such that each $s$-$t$ cut in this area yields a feasible bipartition of the original graph/pair of blocks \textit{within} the balance constraint.
One can then apply a max-flow min-cut algorithm to obtain a min-cut in this area and therefore a non-decreased cut between the original pair of blocks.
This is improved in multiple ways, for example, by iteratively applying the method, searching in larger areas for feasible cuts, and applying most balanced minimum cut heuristics. For more details we refer the reader to \cite{kaffpa}. 

\paragraph*{Multi-try FM.}
The second method for improving a given partition is called multi-try FM.
This local improvement method moves nodes between blocks in order to decrease the cut.
Previous $k$-way methods were initialized with \textit{all} boundary nodes, i.e., all boundary nodes were eligible for movement at the beginning.
Roughly speaking, the multi-try FM algorithm is a $k$-way improvement method that is initialized with a \textit{single} boundary node, thus achieving a more localized search. This is repeated several rounds. The algorithm has a higher chance to escape local optima.  For more details about the multi-try FM algorithm we refer the reader to \cite{kaffpa, dissSchulz}.

\paragraph*{Iterated Multilevel Algorithms.}
KaFFPa extends the concept of \emph{iterated multilevel algorithms}  which
 was introduced for graph partitioning by Walshaw et al. \cite{walshaw2004multilevel}.
The main idea is to iterate the multilevel-scheme using different random seeds for coarsening and uncoarsening. 
Once the graph is partitioned, edges that are between two blocks are not contracted so that the given partition can be used as initial partition on the coarsest level. 
This ensures non-decreased partition quality since the refinement algorithms of KaFFPa guarantee no worsening.
KaFFPa uses F-cycles  as a potentially stronger iterated multilevel algorithm.
The detailed description of F-cycles for the multilevel graph partitioning framework can be found in \cite{kaffpa}. 

\subsection{KaFFPaE}
KaFFPaE (KaFFPaEvolutionary) is a distributed evolutionary algorithm to tackle the graph partitioning problem. 
KaFFPaE uses KaFFPa to provide new effective combine and mutation operators.  
Roughly speaking, to combine two partitions of a population of the algorithm, the coarsening phase of KaFFPa is modified such that no cut edge of either of the input partitions is contracted. This ensures that both input partitions can be used as initial partition on the coarsest level and moreover that  exchanging good parts of solutions can be exchanged effectively. 
Intuitively, the combine operator assembles good parts of solutions into a single partition. 
The combine operation framework is very flexible so that a partition can be combined with an arbitrary domain specific graph clustering.
Moreover, the algorithm is parallelized such that each process has its own population of partitions and independently performs combine and mutation operations.
This is combined with a scalable communication protocol similar to randomized rumor spreading to distribute high quality partitions among the processes.
Overall, the system is able to improve the best known partitioning results for many inputs and also in a short amount of time for graphs of moderate size.

\subsection{KaBaPE}

KaFFPa and KaFFPaE compute partitions of very high quality when some imbalance $\epsilon > 0$ is allowed. 
However, they are not very good for small values of $\epsilon$, in particular for the perfectly balanced case $\epsilon=0$. 
Hence, we developed new techniques for the graph partitioning problem with strict balance constraints, that work well for small values of $\epsilon$ including the perfectly balanced case.
The techniques relax the balance constraint for node movements, but globally maintain balance by combining multiple local searches.  
This is done by reducing the combination problem to finding negative cycles in a directed graph, exploiting the existence of efficient algorithms for this problem.
From a meta-heuristic point of view the proposed algorithms increase the neighborhood of a strictly balanced solution in which local search is able to find better solutions. Moreover, we provide efficient ways to explore this neighborhood.
Experiments indicate that previous algorithms have not been able to find such rather complex movements.
KaBaPE also provides balancing variants of these techniques that are able to make infeasible partitions feasible. 
In contrast to previous algorithms such as Scotch \cite{Scotch}, Jostle~\cite{Walshaw07} and Metis~\cite{karypis1998fast}, KaBaPE is able to \emph{guarantee} that the output partition is feasible. 

\subsection{Specialized Techniques for Social Networks}
We also include specialized techniques for social networks. On social networks matching-based multilevel algorithms have the problem that they can not shrink the graph effectively due to the irregular structure of the graphs, To overcome this, we defined an algorithm that contracts size-constraint clusterings \cite{gpviaclustering14}. 
Here, a fast and cut-based label propagation algorithm was used to compute the clusterings. 
The same algorithm can be used during uncoarsening as a fast and very simple local search algorithms. 
To enable the methods use a preconfiguration of the algorithm that has the word social in its name.
For more details on this particular method we refer the reader to~\cite{gpviaclustering14}. 
Additionally, these algorithms can now also be used within the evolutionary algorithms KaFFPaE/KaBaPE.
This work was joint work with Henning Meyerhenke.
Moreover, in Glantz \etal \cite{conductanceER} we looked at algorithms and edge ratings improving the maximum communication volume of a partition (which are not yet integrated into the system).

\subsection{ParHIP -- Parallel High Quality Partitioning}
A large part of the project are distributed memory parallel algorithms designed for networks having a hierarchical cluster structure such as web graphs or social networks. 
 Unfortunately,   previous parallel graph partitioners originally developed for more regular   mesh-like networks do not work well for complex networks.  Here we address this problem by parallelizing and adapting the \emph{label propagation}   technique originally developed for graph clustering.  By introducing size   constraints, label propagation becomes applicable for both the coarsening and   the refinement phase of multilevel graph partitioning. This way we exploit the hierarchical cluster structure present in many complex networks.
   We obtain very high   quality by applying a highly parallel evolutionary algorithm to the coarsest graph. The resulting system is both more scalable and achieves higher quality   than state-of-the-art systems like ParMetis or PT-Scotch. For large complex   networks the performance differences are very big.  As an example, our algorithm   partitions a web graph with 3.3G edges in 16 \emph{seconds} using 512 cores of a high-performance cluster while producing a   high quality partition -- none of the competing systems can handle this graph on our system.
   For more details, we refer the reader to \cite{DBLP:conf/ipps/MeyerhenkeS015}.

\subsection{Better Process Mapping}
Communication and topology aware process mapping is a powerful approach to reduce communication time in parallel applications with known communication patterns on large, distributed memory systems. We address the problem as a quadratic assignment problem (QAP), and include algorithms to construct initial mappings of processes to processors as well as fast local search algorithms to further improve the mappings. By exploiting assumptions that typically hold for applications and modern supercomputer systems such as sparse communication patterns and hierarchically organized communication systems, we arrive at significantly more powerful algorithms for these special QAPs.  Our multilevel construction algorithms employ perfectly balanced graph partitioning techniques and excessively exploit the given communication system hierarchy. For more details, we refer the reader to~\cite{mappingpaperjesper}. Since v3.00 of KaHIP we also offer a global multisection algorithm that partitions the input network along the hierarchy.

\subsection{Edge Partitioning}
Edge-centric distributed computations have appeared as a recent technique to improve the shortcomings of think-like-a-vertex algorithms on large scale-free networks. In order to increase parallelism on this model, \emph{edge partitioning}---partitioning edges into roughly equally sized blocks---has emerged as an alternative to traditional (node-based) graph partitioning.
We include a fast parallel and sequential split-and-connect graph construction algorithm that yield high-quality edge partitions in a scalable way.
Our technique scales to networks with billions of edges, and runs efficiently on thousands of PEs.
Experiments have shown that our algorithm computes solutions of high quality on large real-world networks~\cite{edgepartitioning2019}.

\subsection{Node Separators}
The node separator problem asks to partition the node set of a graph into three sets $A,B$ and $S$ such that the removal of $S$ disconnects $A$ and $B$. 
We use flow-based and localized local search algorithms withing a multilevel framework to compute node separators~\cite{DBLP:journals/corr/Sanders015}.
A common way to obtain a node separator is the following. 
First, we compute a partition of the graph into two sets $V_1$ and $V_2$.
Clearly, the boundary nodes in $V_1$ would yield a feasible separator and so would the boundary nodes in the opposite block $V_2$. 
Since we are interested in a small separator, we could simply use the smaller set of boundary nodes.

It is worth mentioning that we also provide an extended method which can be used to obtain a $k$-way node separator, \ie $k$ blocks $V_1, \ldots, V_k$ and a set $S$ such that after the removal of the nodes in $S$ there no edge running between the blocks $V_1, \ldots, V_k$.
The method of Pothen \etal \cite{pothen1990partitioning} creates a 2-way node separator from the set of cut edges of a previously computed partition. 
As in the original work, we can use this method in our framework as a post-processing step to compute a node separator from a set of cut edges.
The method computes the smallest node separator that can be found by using a subset of the boundary nodes. 
The main idea is to compute a subset $S$ of the boundary nodes such that each cut edge is incident to at least one of the nodes in $S$. 
Such a set called a \emph{vertex cover}. 
It is easy to see that $S$ is a node separator since the removal of $S$ eliminates all cut edges.  
We can then compute a $k$-way node separator by computing a $k$-partition using KaFFPa and applying the described flow problem between all pairs of blocks that share a non-empty boundary afterwards.
All pair-wise separators together are then be used as a $k$-way separator. 

\subsection{Node Ordering}
Applications such as factorization can be sped up significantly for large sparse matrices by interpreting the matrix as a sparse graph and computing a node ordering that minimizes the so-called fill-in. By applying both new and existing data reduction rules exhaustively before nested dissection, we obtain improved quality and at the same time large improvements in running time on a variety of instances. If METIS is installed, the build script also compiles the fast\_node\_ordering program, which runs reductions before running Metis to compute an ordering. The programs are also available through the library. For details about our algorithm, we refer the reader to \cite{DBLP:journals/corr/abs-2004-11315}.
\subsection{Exact ILP Solvers and ILP Improve}
We provide an ILP as well as an ILP to improve a given partition. We extend the neighborhood of the combination problem for multiple local searches by employing integer linear programming. This enables us to find even more complex combinations and hence to further improve solutions. However, out of the box those the ILPs for the problem typically do not scale to large inputs, in particular because the graph partitioning problem has a very large amount of symmetry -- given a partition of the graph, each permutation of the block IDs gives a solution having the same objective and balance. We define a much smaller graph, called model, and solve the graph partitioning problem on the model to optimality by the integer linear program. Besides other things, this model enables us to use symmetry breaking, which allows us to scale to much larger inputs. In order to compile these program you need to run cmake in the build process as cmake ../ -DCMAKE\_BUILD\_TYPE=Release -DUSE\_ILP=On or run ./compile\_withcmake -DUSE\_ILP=On. For details about the algorithms, we refer the reader to \cite{DBLP:conf/wea/HenzingerN018}.
\section{Graph Format}
\label{ss:graphformat}
\subsection{Input File Format}
\subsubsection{Metis File Format}
The graph format used by our partitioning programs is the same as used by Metis \cite{karypis1998fast}, Chaco \cite{chaco} and the graph format that has been used during the 10th DIMACS Implementation Challenge on Graph Clustering and Partitioning. 
The input graph has to be undirected, without self-loops and without parallel edges.

To give a description of the graph format, we follow the description of the Metis 4.0 user guide very closely. A graph $G=(V,E)$ with $n$ vertices and $m$ edges is stored in a plain text file that contains $n+1$ lines (excluding comment lines). The first line contains information about the size and the type of the graph, while the remaining $n$ lines contain information for each vertex of $G$. Any line that starts with \% is a comment line and is skipped.

The first line in the file contains either two integers, $n$ $m$, or three integers, $n$ $m$ $f$. The first two integers are the number of vertices $n$ and the number of undirected edges of the graph, respectively. Note that in determining the number of edges $m$, an edge between any pair of vertices $v$ and $u$ is counted \emph{only once} and not twice, \ie we do not count the edge $(v,u)$ from $(u,v)$ separately. The third integer $f$ is used to specify whether or not the graph has weights associated with its vertices, its edges or both. If the graph is unweighted then this parameter can be omitted. It should be set to $1$ if the graph has edge weights, 10 if the graph has node weights and 11 if the graph has edge and node weights.

The remaining $n$ lines of the file store information about the actual structure of the graph. In particular, the $i$th line (again excluding comment lines) contains information about the $i$th vertex. Depending on the value of $f$, the information stored in each line is somewhat different. In the most general form (when $f=11$, \ie we have node and edge weights) each line has the following structure:
\begin{center}
       $c\, v_1\, w_1\, v_2\, w_2 \ldots v_k\, w_k$ 
\end{center}
where $c$ is the vertex weight associated with this vertex, $v_1, \ldots, v_k$ are the vertices adjacent to this vertex, and $w_1, \ldots, w_k$ are the weights of the edges. Note that the vertices are numbered starting from 1 (not from 0). Furthermore, the vertex-weights must be integers greater or equal to 0, whereas the edge-weights must be strictly greater than 0.

\begin{figure}[h!]
\begin{center}
\includegraphics[width=.8\textwidth]{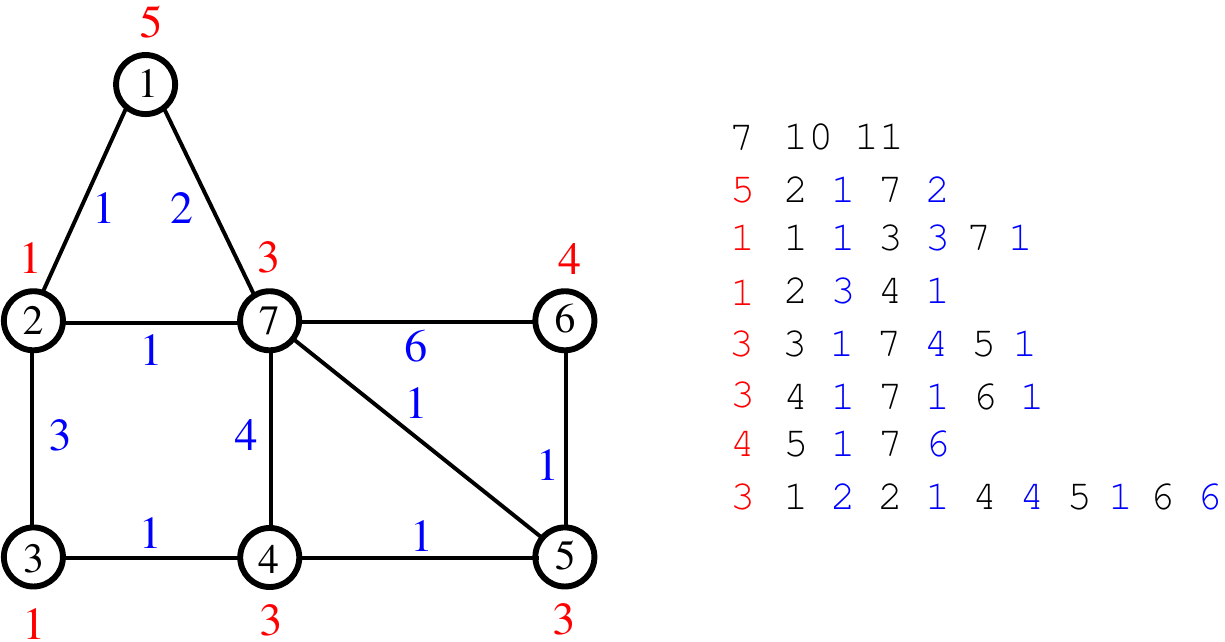}
\end{center}

\caption{An example graph and its representation in the graph format. The IDs of the vertices are drawn within the cycle, the vertex weight is shown next to the circle ({\color{red}red}) and the edge weight is plotted next to the edge ({\color{blue}blue}).}
\label{fig:example}
\end{figure}

\subsubsection{Binary File Format for ParHIP}
In order to read files efficiently in parallel, we also use a binary file format. We give a brief description here, for an in-depth description have a look at parallel\_graph\_io.cpp:483-535). However note that we provide tools to convert the input file format from the previous section into the binary file format. This includes an external memory converter if the graph that needs to be converted does not fit into the memory that is used for conversion. Additionally, ParHIP can also read the simple Metis format (less efficiently). 

The file uses 64bit unsigned longs. It starts with three 64bit unsigned longs, the version number (currently set to 3), the number of nodes $n$ and the third one the number of edges of the graph. The next $n+1$ unsigned longs store offsets. More precisely, the $i$th long is the \emph{position} in which the outgoing edges of vertex $i$ start. Starting at this position the edge targets are stored (one unsigned long for each). The $n+1$ long in the set of offsets stores can be used to determine the end of the file. Node IDs start at 0.
\subsection{Output File Formats}
\subsubsection{Partition}
The output format of a partition is also similar to the output format of a partition provided by Metis. 
It is basically a text file named \emph{tmppartition}$k$ where $k$ is the number of blocks given as input to the partitioning program. 
This file contains $n$ lines. 
In each line the block ID of the corresponding vertex is given, \ie line $i$ contains the block ID of the vertex $i$ (here the vertices are numbered from 0 to $n-1$).
The block IDs are numbered consecutively from 0 to $k-1$. ParHip also has the option to store partitions in a binary format. These can also be converted using the toolbox program.
In case of an edge partition, the file contains $m$ lines in which line $i$ constains the block ID of edge $i$ (here the edges are numbered from 0 to $m-1$).
\subsubsection{Node Separator}
If the output is a node separator then the same format as used for a partition is used. However, in this case the nodes of the separator get the block ID $k$ where as the other nodes maintain their original block id. 
\subsection{Troubleshooting}
KaHIP should not crash! If KaHIP crashes it is mostly due to the following reasons: the provided graph contains self-loops or parallel edges, there exists a forward edge but the backward edge is missing or the forward and backward edges have different weights, or the number of vertices or edges specified does not match the number of vertices or edges provided in the file.
Please use the \emph{graphcheck} tool provided in our graph partitioning package to verify whether your graph has the right input format. If our graphcheck tool tells you that the graph that you provided has the correct format and KaHIP crashes anyway, please write us an email.

\section{User Interface}
KaHIP contains the following programs: kaffpa, kaffpaE, partition\_to\_vertex\_separator, node\_separator, graphchecker, evaluator, parhip, graph2binary, graph2binary\_external, toolbox, node\_ordering, fast\_node\_ordering, edge\_partitioning, distributed\_edge\_partitioning, global\_multisection, ilp\_exact, ilp\_improve. To compile these programs you need to have Argtable, g++, OpenMPI and cmake installed. Once you have that you can execute \emph{compile\_withcmake.sh} in the main folder of the release. For an alternative build process hava a look at the readme file. When the process is finished the binaries can be found in the folder \emph{deploy}. We also provide the option to link against TCMalloc. If you have it installed, run cmake with the additional option -DUSE\_TCMALLOC=On.  By default node ordering programs are also compiled. If you have Metis installed, the build script also compiles a faster node ordering program that uses reductions bevor calling Metis ND.  
If you use the option -DUSE\_ILP=On and you have Gurobi installed, the build script compiles the ILP programs to improve a given partition ilp\_improve and an exact solver ilp\_exact. Alternatively, you can also pass these options to ./compile\_withmake.sh for example: ./compile\_withcmake -DUSE\_ILP=On 
Lastly, we also provide an option to support 64 bit edges. In order to use this, compile KaHIP with the option -D64BITMODE=On. We now explain the parameters of each of the programs briefly.

\subsection*{General Guide:}
There are many programs contained already in the framework. The following table may helps you to find the right program for your use case:
\begin{table}[h]
\begin{tabular}{|l|l|}
\hline
Use case & Programs \\
\hline
\hline
\emph{Default Partitioning Problem} & kaffpa, kaffpaE \\
                \hline
                \hline
Checking Graph for Correctness & graph\_checker \\
Evaluate Partitioning Metrics & evaluator \\
Fast Sequential Partitioning, Mesh  & kaffpa with preconfiguration set to fast \\
Good Sequential Partitioning, Mesh  & kaffpa with preconfiguration set to eco \\
Very Good Sequential Partitioning, Mesh  & kaffpa with preconfiguration set to strong \\
                \hline
Fast Sequential Partitioning, Social  & kaffpa with preconfiguration set to fastsocial \\
Good Sequential Partitioning, Social  & kaffpa with preconfiguration set to ecosocial \\
Very Good Sequential Partitioning, Social & kaffpa with preconfiguration set to strongsocial \\
                \hline
                Mapping to Processor Networks & use the enable\_mapping option \\
                \hline
Highest Quality, Mesh & kaffpaE, use mpirun, large time limit \\
Highest Quality, Social & kaffpaE, use mpirun, large time limit, preconfig strongsocial  \\
        \hline
        \hline
\emph{Parallel Partitioning} & parhip, graph2binary, graph2binary\_external, toolbox\\
                \hline
                \hline
Distributed Memory Parallel, Mesh & parhip, preconfigs ecomesh, fastmesh, ultrafastmesh \\
Distributed Memory Parallel, Social & parhip, preconfigs ecosocial, fastsocial, ultrafastsocial \\
Convert Metis to Binary & graph2binary, graph2binary\_external \\
Evaluate and Convert Partitions & toolbox \\
        \hline
        \hline
\emph{Node Separators} & partition\_to\_vertex\_separator, node\_separator\\
\emph{Node Ordering} & node\_ordering, fast\_node\_ordering\\
                \hline
                \hline
\emph{Edge Partitioning} & edge\_partitioning, distributed\_edge\_partitioning\\
        \hline
        \hline
2-way  Separators & node\_separator\\
$k$-way Separators  & use kaffpa to create $k$-way partition,\\
                    & and then partition\_to\_vertex\_separator to create separator \\
        \hline
        \hline
Process Mapping & kaffpa, with option --enable\_mapping, and global\_multisection program\\
        \hline
        \hline
Exact Solver / Improver & ilp\_exact, ilp\_improve\\        
        \hline
\end{tabular}
\end{table}

\vfill\pagebreak
\ \pagebreak
\subsection{KaFFPa}
\paragraph*{Description:} This is the multilevel graph partitioning program. 
\paragraph*{Usage:\\}

\begin{tabular}{ll}
kaffpa &   file -{}-k=<int> [-{}-help] [-{}-seed=<int>]  [-{}-preconfiguration=variant] [-{}-imbalance=<double>] \\
       &  [-{}-time\_limit=<double>] [-{}-enforce\_balance] [-{}-input\_partition=<string>] \\
       & [-{}-balance\_edges] [-{}-output\_filename=<string>] \\
       & [--enable\_mapping -{}-hierarchy\_parameter\_string=<string> \\
       & -{}-distance\_parameter\_string=<string> [-{}-online\_distances]]
\end{tabular}
\subsection*{Options:\\}

\begin{tabularx}{\textwidth}{lX}
  file                        & Path to graph file that you want to partition. \\
  -{}-k=<int>                   & Number of blocks to partition the graph into. \\
  -{}-help                      & Print help. \\
  -{}-seed=<int>                & Seed to use for the random number generator. \\
  -{}-preconfiguration=variant & Use a preconfiguration. (Default: eco) [strong| eco | fast | fastsocial| ecosocial| strongsocial ]. Strong should be used if quality is paramount, eco if you need a good tradeoff between partition quality and running time, and fast if partitioning speed is in your focus. Configurations with a social in their name should be used for social networks and web graphs. \\
  -{}-imbalance=<double>        & Desired balance. Default: 3 (\%). \\
  -{}-time\_limit=<double>      & Time limit in seconds s. The default value is set to 0s, \ie one partitioner call will be made. If you set a time limit $t$, kaffpa will repeatedly call the multilevel method until the time limit is reached and return the best solution found. \\
  -{}-enforce\_balance          & Use this option only on graphs without vertex weights. If this option is enabled, kaffpa guarantees that the output partition is feasible, \ie fulfills the specified balance constraint. \\
  -{}-balance\_edges          & Use this option to balance the edges among the blocks as well as the nodes. In this case node weights are set to $c(v)+\text{deg}_\omega(v)$ for a node $v\in V$ and the balance constraint is adapted accordingly. \\
  -{}-input\_partition=<string> & You can specify an input partition. If you do so, KaFFPa will try to improve it.   \\
  -{}-enable\_mapping            & Enable mapping algorithms to map quotient graph onto processor graph defined by hierarchy and distance options. (Default: disabled) \\
  -{}-hierarchy\_parameter\_string=<string>    &Specify as 4:8:8 for 4 cores per PE, 8 PEs per rack, ... and so forth. \\
  -{}-distance\_parameter\_string=<string>     &Specify as 1:10:100 if cores on the same chip have distance 1, PEs in the same rack have distance 10, ... and so forth. \\
  -{}-online\_distances                       & Do not store processor distances in a matrix, but do recomputation. (Default: disabled) \\
  -{}-output\_filename=<string>               & Specify the output filename (default tmppartition\$k). \\
\end{tabularx}
\vfill
\pagebreak
\subsection{KaFFPaE / KaBaPE}
\paragraph*{Description:} This is the distributed evolutionary algorithm to tackle the graph partitioning problem. It includes also the perfectly balance case $\epsilon=0$.
\paragraph*{Usage:\\} 
\begin{tabular}{ll}
mpirun -n P kaffpaE & file -{}-k=<int> [-{}-help] [-{}-seed=<int>]  [-{}-preconfiguration=variant] [-{}-imbalance=<double>] \\
                    & [-{}-time\_limit=<double>] [-{}-mh\_enable\_quickstart] [-{}-mh\_optimize\_communication\_volume]\\
                    & [-{}-mh\_enable\_kabapE] [-{}-mh\_enable\_tabu\_search] [-{}-kabaE\_internal\_bal=<double>]\\ 
                    & [-{}-input\_partition=<string>] [-{}-balance\_edges] [-{}-output\_filename=<string> ]
\end{tabular}

\subsection*{Options:\\} 
\begin{tabularx}{\textwidth}{lX}
  P                                     & Number of processes to use.  \\
  file                        & Path to graph file that you want to partition. \\
  -{}-k=<int>                             & Number of blocks to partition the graph into.\\
  -{}-help                                & Print help. \\
  -{}-seed=<int>                          & Seed to use for the random number generator. \\
  -{}-preconfiguration=variant & Use a preconfiguration. (Default: strong) [strong | eco | fast | fastsocial | ecosocial | strongsocial ]. Strong should be used if quality is paramount, eco if you need a good tradeoff between partition quality and running time, and fast if partitioning speed is in your focus. Configurations with a social in their name should be used for social networks and web graphs.\\
  -{}-imbalance=<double>                     & Desired balance. Default: 3 (\%).\\
  -{}-time\_limit=<double>                & Time limit in seconds s. The default value is set to 0s, \ie one partitioner call will be made by each PE. In order to use combine operations you \emph{have to} set a time limit $t>0$. kaffpaE will return the best solution after the time limit is reached. A time limit $t=0$ means that the algorithm will only create the initial population. \\
  -{}-mh\_enable\_quickstart              & Enables the quickstart option. In this case each PE creates a few partitions and these partitions are distributed among the PEs.\\
  -{}-mh\_optimize\_communication\_volume & Modifies the fitness function of the evolutionary algorithm so that communication volume is optimized.\\
  -{}-mh\_enable\_kabapE                  & Enables the combine operator of \emph{KaBaPE}.\\
  -{}-mh\_enable\_tabu\_search            & Enables our version of combine operation by block matching.\\
  -{}-kabaE\_internal\_bal=<double>       & The internal balance parameter for KaBaPE (Default: 0.01) (1 \%)\\
  -{}-balance\_edges                      & Use this option to balance the edges among the blocks as well as the nodes. In this case node weights are set to $c(v)+\text{deg}_\omega(v)$ for a node $v\in V$ and the balance constraint is adapted accordingly. \\
  -{}-input\_partition=<string> & You can specify an input partition. If you do so, KaFFPaE will try to improve it.  \\
  -{}-output\_filename=<string>               & Specify the output filename (default tmppartition\$k). \\
\end{tabularx}
\vfill
\pagebreak
\subsection{ParHIP -- Parallel High Quality Partitioning}
\subsubsection{ParHIP}
\paragraph*{Description:} This is the distributed memory parallel program to compute a partition. 
\paragraph*{Usage:\\} 
\begin{tabular}{ll}
mpirun -n P parhip & file -{}-k=<int> [-{}-help] [-{}-seed=<int>]  [-{}-preconfiguration=variant] [-{}-imbalance=<int>] \\
                    & [-{}-vertex\_degree\_weights] [-{}-save\_partition] [-{}-save\_partition\_binary ]
\end{tabular}

\subsubsection*{Options:\\} 
\begin{tabularx}{\textwidth}{lX}
  P                                     & Number of processes to use.  \\
  file                                  &   Path to graph file to partition. (Either Metis format or binary format)\\
  -{}-seed=<int>                           &  Seed to use for the PRNG. \\
  -{}-k=<int>                               & Number of blocks to partition the graph.\\
  -{}-help                                 &  Print help.\\
  -{}-imbalance=<int>                      &  Desired balance. Default: 3 (\%).\\
  -{}-preconfiguration=variant &  Use a preconfiguration. (Default: fast) [ecosocial | fastsocial | ultrafastsocial | ecomesh | fastmesh | ultrafastmesh].\\
  -{}-vertex\_degree\_weights                &  Use 1+deg(v) as vertex weights.\\
  -{}-save\_partition                       &  Enable this tag if you want to store the partition to disk.\\
  -{}-save\_partition\_binary                &  Enable this tag if you want to store the partition to disk in a binary format.\\
\end{tabularx}
\subsubsection{Graph Conversion}
\paragraph*{Description:} This is the program to convert a Metis graph format into the binary graph format. The program ending with \_external does the task in external memory.
\paragraph*{Usage:\\} 
\begin{tabular}{ll}
graph2binary[\_external] & metisfile outputfilename
\end{tabular}
\vfill\pagebreak
\subsubsection{Evaluation}
\paragraph*{Description:} This is the toolbox program to convert partitions and to evaluate them.
\paragraph*{Usage:\\} 
\begin{tabular}{ll}
mpirun -n P toolbox & file -{}-k=<int> [-{}-help] =<int>]  -{}-input\_partition=<string>  \\
                    & [-{}-save\_partition] [-{}-save\_partition\_binary ] [-{}-evaluate]
\end{tabular}
\subsection*{Options:\\} 
\begin{tabularx}{\textwidth}{lX}
  P                                     & Number of processes to use.  \\
  file                                  &   Path to graph file to partition. (Either Metis format or binary format)\\
  -{}-k=<int>                               & Number of blocks to graph is partitioned in.\\
  -{}-help                                 &  Print help.\\
  -{}-input\_partition=file                 &  Path to partition file to convert. \\
  -{}-save\_partition                       &  Enable this tag if you want to store the partition to disk.\\
  -{}-save\_partition\_binary                &  Enable this tag if you want to store the partition to disk in a binary format.\\
  -{}-evaluate                             &  Enable this tag the partition to be evaluated.

\end{tabularx}

\vfill
\pagebreak
\subsection{Node Separators}
\subsubsection{Partition Converter}
\paragraph*{Description:} This is the program that computes a $k$-way vertex separator  given a $k$-way partition of the graph. Use this approach if $k>2$.
\paragraph*{Usage:\\} 
\begin{tabular}{ll}
partition\_to\_vertex\_separator & file -{}-k=<int> -{}-input\_partition=<string> [-{}-help] [-{}-seed=<int>] [-{}-output\_filename]
\end{tabular}
\subsection*{Options:\\} 
\begin{tabularx}{\textwidth}{lX}
  file                       & Path to the graph file. \\
  -{}-k=<int>                  & Number of blocks the graph is partitioned in by using the input partition. \\
  -{}-input\_partition=<string>& Input partition to compute the vertex separator from. \\
  -{}-help                     & Print help. \\
  -{}-seed=<int>               & Seed to use for the random number generator. \\
  -{}-output\_filename=<string>               & Specify the output filename (default tmpseparator). \\
\end{tabularx}
\subsubsection{Biseparators}
\paragraph*{Description:} This is the program that computes a $2$-way vertex separator. Use this approach if $k=2$.
\paragraph*{Usage:\\} 
\begin{tabular}{ll}
node\_separator & file [-{}-seed=<int>]  [-{}-preconfiguration=variant]  [-{}-help] [-{}-output\_filename] 
\end{tabular}
\subsection*{Options:\\} 
\begin{tabularx}{\textwidth}{lX}
  file                       & Path to the graph file. \\
  -{}-help                     & Print help. \\
  -{}-imbalance=<double>                     & Desired balance. Default: 20 (\%).\\
  -{}-seed=<int>               & Seed to use for the random number generator. \\
  -{}-preconfiguration=variant & Use a preconfiguration. (Default: strong) [strong | eco | fast | fastsocial | ecosocial | strongsocial ]. Strong should be used if quality is paramount, eco if you need a good tradeoff between partition quality and running time, and fast if partitioning speed is in your focus. Configurations with a social in their name should be used for social networks and web graphs (they use a different kind of coarsening scheme).\\
  -{}-output\_filename=<string>               & Specify the output filename (default tmpseparator). \\
\end{tabularx}
\vfill
\pagebreak
\subsection{Edge Partitioning}
\paragraph*{Description:} This is the edge partitioning program. 
\paragraph*{Usage:\\}

\begin{tabular}{ll}
edge\_partitioning&   file -{}-k=<int> [-{}-help] [-{}-seed=<int>]  -{}-preconfiguration=variant [-{}-imbalance=<double>] \\
       & [-{}-infinity=<int>] [-{}-output\_filename=<string>]
\end{tabular}
\subsection*{Options:\\}

\begin{tabularx}{\textwidth}{lX}
  file                        & Path to graph file that you want to partition. \\
  -{}-k=<int>                   & Number of blocks to partition the graph into. \\
  -{}-help                      & Print help. \\
  -{}-seed=<int>                & Seed to use for the random number generator. \\
  -{}-preconfiguration=variant & Use a preconfiguration. (Default: eco) [strong| eco | fast | fastsocial| ecosocial| strongsocial ]. Strong should be used if quality is paramount, eco if you need a good tradeoff between partition quality and running time, and fast if partitioning speed is in your focus. Configurations with a social in their name should be used for social networks and web graphs. \\
  -{}-imbalance=<double>        & Desired balance. Default: 3 (\%). \\
  -{}-infinity=<int>               & Infinity edge weight used in the SPAC model. Default: 1000. \\
  -{}-output\_filename=<string>               & Specify the output filename (default tmpedgepartition\$k). \\
\end{tabularx}
\subsection{Distributed Edge Partitioning}
\paragraph*{Description:} This is the distributed memory parallel edge partitioning program. 
\paragraph*{Usage:\\}

\begin{tabular}{ll}
mpirun -n N distributed\_edge\_partitioning&   file -{}-k=<int> [-{}-help] [-{}-seed=<int>]  -{}-preconfiguration=variant  \\
       & [-{}-infinity=<int>] [-{}-output\_filename=<string>] [-{}-imbalance=<double>]
\end{tabular}
\subsection*{Options:\\}

\begin{tabularx}{\textwidth}{lX}
  file                        & Path to graph file that you want to partition. \\
  -{}-k=<int>                   & Number of blocks to partition the graph into. \\
  -{}-help                      & Print help. \\
  -{}-seed=<int>                & Seed to use for the random number generator. \\
  -{}-preconfiguration=variant & Use a preconfiguration. (Default: eco) [strong| eco | fast | fastsocial| ecosocial| strongsocial ]. Strong should be used if quality is paramount, eco if you need a good tradeoff between partition quality and running time, and fast if partitioning speed is in your focus. Configurations with a social in their name should be used for social networks and web graphs. \\
  -{}-imbalance=<double>        & Desired balance. Default: 3 (\%). \\
  -{}-infinity=<int>               & Infinity edge weight used in the SPAC model. Default: 1000000. \\
  -{}-save\_partition                       &  Enable this tag if you want to store the partition to disk.\\
  -{}-save\_partition\_binary                &  Enable this tag if you want to store the partition to disk in a binary format.\\
\end{tabularx}

\vfill
\pagebreak
\subsection{Node Ordering}
\paragraph*{Description:} This is node ordering program. 
\paragraph*{Usage:\\}

\begin{tabular}{ll}
node\_ordering &   file  [-{}-help] FILE [-{}-seed=<int>] [-{}-output\_filename=<string>] \\ &[-{}-preconfiguration=VARIANT] [-{}-reduction\_order=<string>] \\
fast\_node\_ordering &   file  [-{}-help] FILE [-{}-seed=<int>] [-{}-output\_filename=<string>] \\ &[-{}-preconfiguration=VARIANT] [-{}-reduction\_order=<string>] \\

\end{tabular}
\subsection*{Options:\\}

\begin{tabularx}{\textwidth}{lX}
  file                        & Path to graph file that you want to partition. \\
  -{}-help                      & Print help. \\
  -{}-seed=<int>                & Seed to use for the random number generator. \\
  -{}-preconfiguration=variant & Use a preconfiguration. (Default: eco) [strong| eco | fast | fastsocial| ecosocial| strongsocial ]. Strong should be used if quality is paramount, eco if you need a good tradeoff between ordering quality and running time, and fast if ordering speed is in your focus. Configurations with a social in their name should be used for social networks and web graphs. \\
  -{}-output\_filename=<string>               & Specify the output filename (default tmppartition\$k). \\
-{}-reduction\_order=<string>    &           Order in which to apply reductions. Reduction numbers 0-5. Specify as string, for example "0 4". Available reductions: 0 simplical\ node\ reduction, 1 indistinguishable\_nodes, 2 twins, 3 path\_compression, 4 degree\_2\_nodes, 5 triangle\_contraction.
\end{tabularx}
\pagebreak
\subsection{Process Mapping}
\paragraph*{Description:} This is the multilevel process mapping program. Note that $k$ is given implicitely by the specification of the hierarchy.
\paragraph*{Usage:\\} 
\begin{tabular}{ll}
global\_multisection &   file [-{}-help] [-{}-seed=<int>]  [-{}-preconfiguration=variant] [-{}-imbalance=<double>] \\
       &  [-{}-time\_limit=<double>] [-{}-enforce\_balance] [-{}-input\_partition=<string>] \\
       &  [-{}-output\_filename=<string>] \\
       &  -{}-hierarchy\_parameter\_string=<string> \\
       & -{}-distance\_parameter\_string=<string> [-{}-online\_distances]
\end{tabular}
\subsection*{Options:\\}

\begin{tabularx}{\textwidth}{lX}
  file                        & Path to graph file that you want to partition. \\
  -{}-help                      & Print help. \\
  -{}-seed=<int>                & Seed to use for the random number generator. \\
  -{}-preconfiguration=variant & Use a preconfiguration. (Default: eco) [strong| eco | fast | fastsocial| ecosocial| strongsocial ]. Strong should be used if quality is paramount, eco if you need a good tradeoff between partition quality and running time, and fast if partitioning speed is in your focus. Configurations with a social in their name should be used for social networks and web graphs. \\
  -{}-imbalance=<double>        & Desired balance. Default: 3 (\%). \\
  -{}-time\_limit=<double>      & Time limit in seconds s. The default value is set to 0s, \ie one partitioner call will be made. If you set a time limit $t$, kaffpa will repeatedly call the multilevel method until the time limit is reached and return the best solution found. \\
  -{}-enforce\_balance          & Use this option only on graphs without vertex weights. If this option is enabled, kaffpa guarantees that the output partition is feasible, \ie fulfills the specified balance constraint. \\
  -{}-input\_partition=<string> & You can specify an input partition. If you do so, KaFFPa will try to improve it.   \\
  -{}-hierarchy\_parameter\_string=<string>    &Specify as 4:8:8 for 4 cores per PE, 8 PEs per rack, ... and so forth. \\
  -{}-distance\_parameter\_string=<string>     &Specify as 1:10:100 if cores on the same chip have distance 1, PEs in the same rack have distance 10, ... and so forth. \\
  -{}-online\_distances                       & Do not store processor distances in a matrix, but do recomputation. (Default: disabled) \\
  -{}-output\_filename=<string>               & Specify the output filename (default tmppartition\$k). \\
\end{tabularx}

\pagebreak
\subsection{ILP Exact Solver / Improvement}
\paragraph*{Description:} This is the ILP program solve the problem to optimality.
\paragraph*{Usage:\\}

\begin{tabular}{ll}
ilp\_exact & [-{}-help] FILE [-{}-seed=<int>] -{}-k=<int> [-{}-output\_filename=<string>] [-{}-imbalance=<double>] \\&[-{}-ilp\_timeout=<int>] \\
\end{tabular}
\subsection*{Options:\\}

\begin{tabularx}{\textwidth}{lX}
  file                        & Path to graph file that you want to partition. \\
  -{}-k=<int>                   & Number of blocks to partition the graph into. \\
  -{}-help                      & Print help. \\
  -{}-seed=<int>                & Seed to use for the random number generator. \\
  -{}-ilp\_timeout=<int>       &                ILP timeout in seconds (Default: 7200) \\
  -{}-imbalance=<double>        & Desired balance. Default: 3 (\%). \\
  -{}-output\_filename=<string>               & Specify the output filename (default tmppartition\$k). \\
\end{tabularx}
\subsubsection{Improvements via ILPs}
\paragraph*{Description:} This are the ILP programs to improve a partition.
\paragraph*{Usage:\\}

\begin{tabular}{ll}
ilp\_improve & [-{}-help] FILE [-{}-seed=<int>] -{}-k=<int> [-{}-output\_filename=<string>] [-{}-imbalance=<double>] \\&[-{}-ilp\_timeout=<int>] -{}-input\_partition=<string>  [-{}-ilp\_mode=<string>] [-{}-ilp\_min\_gain=<int>] \\& [-{}-ilp\_bfs\_depth=<int>] [-{}-ilp\_overlap\_presets=<string>] [-{}-ilp\_limit\_nonzeroes=<int>]\\& [-{}-ilp\_overlap\_runs=<int>] [-{}-ilp\_timeout=<int>]\\
\end{tabular}
\subsection*{Options:\\}

\begin{tabularx}{\textwidth}{lX}
  file                        & Path to graph file that you want to partition. \\
  -{}-k=<int>                   & Number of blocks to partition the graph into. \\
  -{}-help                      & Print help. \\
  -{}-seed=<int>                & Seed to use for the random number generator. \\
  -{}-ilp\_timeout=<int>       &                ILP timeout in seconds (Default: 7200) \\
  -{}-imbalance=<double>        & Desired balance. Default: 3 (\%). \\
  -{}-output\_filename=<string>               & Specify the output filename (default tmppartition\$k). \\
  -{}-input\_partition=file                 &  Path to partition file to convert. \\
-{}-ilp\_mode=<string>        &              ILP Localsearch mode [boundary|gain|trees|overlap]. \\
  -{}-ilp\_min\_gain=<int>     &                In Gain mode: build BFS around each vertex with gain >= ilp\_min\_gain [default: -1]\\
  -{}-ilp\_bfs\_depth=<int>     &               Depth of BFS trees in ILP improve [default: 2]\\
  -{}-ilp\_overlap\_presets=<string>          & In overlap mode: fix assignment of vertices to break symmetries [none,random,noequal,center,heaviest] (Default: noequal)\\
  -{}-ilp\_limit\_nonzeroes=<int>             & ILP: Limit of nonzeroes in ILP problem. [default: 5,000,000]\\
  -{}-ilp\_overlap\_runs=<int>                & In overlap mode: Build overlap using ilp\_overlap\_runs many subproblem.\\
\end{tabularx}

\subsection{Label Propagation}
\paragraph*{Description:} This is the program that a size-constrained label propagation clustering.
\paragraph*{Usage:\\} 
\begin{tabular}{ll}
label\_propagation & file [-{}-cluster\_upperbound=<int>] [-{}-label\_propagation\_iterations] [-{}-help] [-{}-seed=<int>]\\
               & [-{}-output\_filename]
\end{tabular}
\subsection*{Options:\\} 
\begin{tabularx}{\textwidth}{lX}
  file                       & Path to the graph file. \\
  -{}-cluster\_upperbound=<int>             &  Set a size-constraint on the size of a cluster. For example, specifying a value of 10 means that each cluster can have at most 10 vertices (or a weight of 10 if the graph is weighted). By default there is no size-constraint, so that each node can potentially be in on cluster.  \\
  -{}-label\_propagation\_iterations=<int>  &   Set the number of label propagation iterations. The default value is 10. \\

  -{}-help                     & Print help. \\
  -{}-seed=<int>               & Seed to use for the random number generator. \\
  -{}-output\_filename=<string>               & Specify the output filename (default tmpclustering). \\
\end{tabularx}

\subsection{Graph Format Checker}
\paragraph*{Description:} This program checks if the graph specified in a given file is valid. 
\paragraph*{Usage:\\} 
\begin{tabular}{ll}
graphchecker & file
\end{tabular}
\subsection*{Options:\\} 
\begin{tabularx}{\textwidth}{lX}
  file                       & Path to the graph file. \\
\end{tabularx}

\vfill
\pagebreak
\section{Library}
Some of the programs above can be directly accessed in C/C++ by using the library that we provide, \ie the graph partitioners KaFFPaFast, KaFFPaEco and KaFFPaStrong (more will follow) can be called using this library. In this section we describe how the library can be used, the methods can be accessed and describe the parameters of these functions. The functions described in this section can be found in the file \emph{interface/kaHIP\_interface.h}. \textbf{An example program that calls and links the library can be found in \emph{misc/example\_library\_call/}.} This is also compiled during the build process. More over the library also contains functions for node ordering, separators as well as process mapping functionality.

\subsection{Graph Data Structure}
\label{ss:gds}
To make it easy for you to try KaHIP, we use a more or less similar interface to that provided by Metis. 
Most importantly, we use the same graph data structure. 
The graph data structure is an adjacency structure of the graph which includes weights for the vertices and edges if there are any. 
The adjacency structure is a compressed sparse row (CSR) format. 
We closely follow the description of the Metis 4.0 user guide: 
\begin{figure}[t!]
\begin{center}
        
\hspace*{4cm}
\begin{tabularx}{\textwidth}{lX}
        \includegraphics[width=2.75cm]{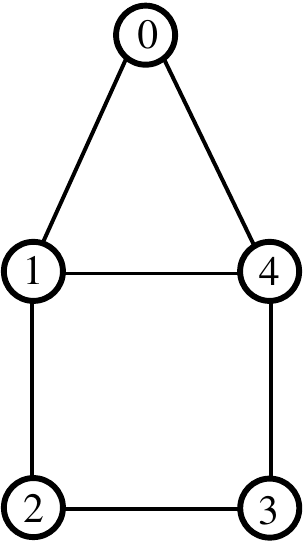} & \vspace*{-3.5cm}\begin{minipage}{5cm}\emph{xadj}: \hspace*{.35cm}0 2 5 7 9 12 \\ 
                                                                 \emph{adjncy}: 1 4 0 2 4 1 3 2 4 0 1 3\end{minipage}

\end{tabularx}
\end{center}

\caption{An example unweighted graph with its graph data structure.}
\label{fig:}
\end{figure}

The CSR format is a widely used scheme for storing sparse graphs. In this format the adjacency structure of a graph with $n$ vertices and $m$ edges is represented using two arrays \emph{xadj} and \emph{adjncy}. The \emph{xadj} array is of size $n+1$ whereas the \emph{adjncy} array is of size $2m$ (this is because for each edge between vertices $v$ and $u$ we store both a forward edge $(v,u)$ and a backward edge $(u,v)$). 

The adjacency structure of the graph is stored as follows. First of all, the vertex numbering starts from 0. The adjacency list of vertex $i$ is stored in the array \emph{adjncy} starting at index \emph{xadj}[$i$] and ending at (but not including) index \emph{xadj}[$i+1$] (\ie \emph{adjncy}[\emph{xadj}[$i$]] through and including \emph{adjncy}[\emph{xadj}[$i+1$]$-1$]).
That is, for each vertex $i$, its adjacency list is stored in consecutive locations in the array \emph{adjncy}, and the array \emph{xadj} is used to point to where it begins and where it ends. 

The weights of the vertices (if any) are stored in an additional array called \emph{vwgt}. If the graph has $n$ vertices then the array contains $n$ elements, and \emph{vwgt}[$i$] will store the weight of the $i$th vertex. The vertex-weights must be integers greater or equal to zero. If all the vertices of the graph have the same weight (\eg the graph is unweighted), then the array \emph{vwgt} can be set to NULL.

The weights of the edges (if any) are stored in additional array called adjwgt. This array contains $2m$ elements, and the weight of edge \emph{adjncy}[j] is stored at location \emph{adjwgt}[j]. The edge-weights must be integers greater than zero. 
The weight of a forward edge $(u,v)$ has to be equal to the weight of the backward edge $(v,u)$ and parallel edges as well as self-loops are not allowed.
If all the edges of the graph have the same weight (\eg the graph is unweighted), then the array \emph{adjwgt} can be set to NULL.

\subsection{Functions}
Currently we provide functions for partitioning, node\_separators, node ordering, as well as process mapping. The function kaffpa corresponds to the mulilevel partitioner KaFFPa, kaffpa\_balance\_NE also balances the edges among the blocks and the function node\_separator used KaFFPa to compute a partition and from that a node separator is derived, process\_mapping runs process mapping algorithms, reduced\_nd and reduced\_nd\_fast are the node ordering functions.

\subsubsection*{Main Partitioner Call}
{\color{blue}void} kaffpa({\color{blue}int}* $n$, {\color{blue}int}* \emph{vwgt}, {\color{blue}int}* \emph{xadj}, {\color{blue}int}* \emph{adjcwgt}, {\color{blue}int}* \emph{adjncy},\\ 
\hspace*{2.925cm} {\color{blue}int}* \emph{nparts}, {\color{blue}double}* \emph{imbalance},  {\color{blue}bool} \emph{suppress\_output}, {\color{blue}int} \emph{seed}, {\color{blue} int} \emph{mode},\\
 \hspace*{3cm}{\color{blue}int}* \emph{edgecut}, {\color{blue}int}* \emph{part}); \\
        
\subsubsection*{Parameters}
\begin{tabularx}{\textwidth}{lX}
$n$ & Number of vertices of the graph.\\
\emph{vwgt} & This is the vertex weight array described in Section~\ref{ss:gds}. The array should have size $n$. If your graph does not have vertex weight you can use a null pointer as an argument.\\
\emph{xadj} & This is the xadj array described in Section~\ref{ss:gds} which holds the pointers to the adjacency lists of the vertices. The array should have size $n+1$.\\
\emph{adjcwgt} & This is the adjacency weight array described in Section~\ref{ss:gds} which holds the weights of the edges if they exists. The array should have size $2m$. If your graph does not have edge weights you can use a null pointer as an argument.\\
\emph{adjncy} & This is the adjacency array described in Section~\ref{ss:gds} which holds the adjacency lists of the vertices. The array should have size $2m$.\\
\emph{nparts} & This parameter specifies the number of blocks you want the graph to be partitioned in.\\
\emph{imbalance} & This parameter controls the amount of imbalance that is allowed. For example, setting it to 0.03 specifies an imbalance of 3\% which means on unweighted graphs that each block has to fulfill the constraint $|V_i| \leq (1+0.03)|V|/k$.\\
\emph{suppress\_output} & If this option is enabled then no output of the partitioning library is printed to stdout.\\
\emph{seed} & Seed to use for the random number generator.\\
\emph{mode} & One out of FAST, ECO, STRONG, FASTSOCIAL, ECOSOCIAL, STRONGSOCIAL. Configuration names correpond to the default configuration names of the multilevel partitioner KaFFPa.\\
\emph{edgecut} & This is an \emph{output} parameter. It represents the edge cut of the computed partition.  \\
\emph{part} & This is an \emph{output} parameter. It has to be an already allocated array of size $n$. After the function call this array contains the information of the blocks of the vertices, \ie the block of the $i$th vertex is given in \emph{part}[$i$].  \\
\end{tabularx}
\vfill
\pagebreak
\subsubsection*{Node+Edge Balanced Partitioner Call}
This function will try to balance nodes \emph{and} edges among the blocks. Use this function if you want to use KaFFPa for applications in which workload is also associated with edges, e.g. sparse matrix vector multiplications or in graph processing frameworks. \\ 

\noindent{\color{blue}void} kaffpa\_balance\_NE({\color{blue}int}* $n$, {\color{blue}int}* \emph{vwgt}, {\color{blue}int}* \emph{xadj}, {\color{blue}int}* \emph{adjcwgt}, {\color{blue}int}* \emph{adjncy},\\ 
\hspace*{2.925cm} {\color{blue}int}* \emph{nparts}, {\color{blue}double}* \emph{imbalance},  {\color{blue}bool} \emph{suppress\_output}, {\color{blue}int} \emph{seed}, {\color{blue} int} \emph{mode},\\
 \hspace*{3cm}{\color{blue}int}* \emph{edgecut}, {\color{blue}int}* \emph{part}); \\
        
\subsubsection*{Parameters}
\begin{tabularx}{\textwidth}{lX}
$n$ & Number of vertices of the graph.\\
\emph{vwgt} & This is the vertex weight array described in Section~\ref{ss:gds}. The array should have size $n$. If your graph does not have vertex weight you can use a null pointer as an argument.\\
\emph{xadj} & This is the xadj array described in Section~\ref{ss:gds} which holds the pointers to the adjacency lists of the vertices. The array should have size $n+1$.\\
\emph{adjcwgt} & This is the adjacency weight array described in Section~\ref{ss:gds} which holds the weights of the edges if they exists. The array should have size $2m$. If your graph does not have edge weights you can use a null pointer as an argument.\\
\emph{adjncy} & This is the adjacency array described in Section~\ref{ss:gds} which holds the adjacency lists of the vertices. The array should have size $2m$.\\
\emph{nparts} & This parameter specifies the number of blocks you want the graph to be partitioned in.\\
\emph{imbalance} & This parameter controls the amount of imbalance that is allowed. For example, setting it to 0.03 specifies an imbalance of 3\% which means on unweighted graphs that each block has to fulfill the constraint $|V_i| \leq (1+0.03)|V|/k$.\\
\emph{suppress\_output} & If this option is enabled then no output of the partitioning library is printed to stdout.\\
\emph{seed} & Seed to use for the random number generator.\\
\emph{mode} & One out of FAST, ECO, STRONG, FASTSOCIAL, ECOSOCIAL, STRONGSOCIAL. Configuration names correpond to the default configuration names of the multilevel partitioner KaFFPa.\\
\emph{edgecut} & This is an \emph{output} parameter. It represents the edge cut of the computed partition.  \\
\emph{part} & This is an \emph{output} parameter. It has to be an already allocated array of size $n$. After the function call this array contains the information of the blocks of the vertices, \ie the block of the $i$th vertex is given in \emph{part}[$i$].  \\
\end{tabularx}
\vfill
\pagebreak

\subsubsection*{Node Separator}
{\color{blue}void} node\_separator({\color{blue}int}* $n$, {\color{blue}int}* \emph{vwgt}, {\color{blue}int}* \emph{xadj}, {\color{blue}int}* \emph{adjcwgt}, {\color{blue}int}* \emph{adjncy},\\ 
\hspace*{2.925cm} {\color{blue}int}* \emph{nparts}, {\color{blue}double}* \emph{imbalance},  {\color{blue}bool} \emph{suppress\_output}, {\color{blue}int} \emph{seed}, {\color{blue} int} \emph{mode},\\
 \hspace*{3cm}{\color{blue}int}* \emph{num\_separator\_vertices}, {\color{blue}int}** \emph{separator}); \\

\subsubsection*{Parameters}
\begin{tabularx}{\textwidth}{lX}
$n$ & Number of vertices of the graph.\\
\emph{vwgt} & This is the vertex weight array described in Section~\ref{ss:gds}. The array should have size $n$. If your graph does not have vertex weight you can use a null pointer as an argument.\\
\emph{xadj} & This is the xadj array described in Section~\ref{ss:gds} which holds the pointers to the adjacency lists of the vertices. The array should have size $n+1$.\\
\emph{adjcwgt} & This is the adjacency weight array described in Section~\ref{ss:gds} which holds the weights of the edges if they exists. The array should have size $2m$. If your graph does not have edge weights you can use a null pointer as an argument.\\
\emph{adjncy} & This is the adjacency array described in Section~\ref{ss:gds} which holds the adjacency lists of the vertices. The array should have size $2m$.\\
\emph{nparts} & This parameter specifies the number of blocks you want the graph to be partitioned in as a base to compute a separator from. If the size of the separator is your objective, we recommend to set this parameter to 2.\\
\emph{imbalance} & This parameter controls the amount of imbalance that is allowed. For example, setting it to 0.03 specifies an imbalance of 3\% which means on unweighted graphs that each block has to fulfill the constraint $|V_i| \leq (1+0.03)|V|/k$.\\
\emph{suppress\_output} & If this option is enabled then no output of the partitioning library is printed to stdout.\\
\emph{seed} & Seed to use for the random number generator.\\
\emph{mode} & One out of FAST, ECO, STRONG, FASTSOCIAL, ECOSOCIAL, STRONGSOCIAL. Configuration names correpond to the default configuration names of the multilevel partitioner KaFFPa.\\
\emph{num\_separator\_vertices} & This is an \emph{output} parameter. It represents the number of separator vertices, i.e. the size of the array separator.  \\
\emph{separator} & This is an \emph{output} parameter. After the function call this array contains the ids of the separator vertices.  \\
\end{tabularx}

\vfill
\pagebreak
\subsubsection*{Node Ordering}
{\color{blue}void} reduced\_nd({\color{blue}int}* $n$, {\color{blue}int}* \emph{xadj}, {\color{blue}int}* \emph{adjncy}, {\color{blue}bool} \emph{suppress\_output}, {\color{blue}int} \emph{seed}, {\color{blue} int} \emph{mode}, {\color{blue}int}** \emph{ordering}); \\
        
\noindent {\color{blue}void} fast\_reduced\_nd({\color{blue}int}* $n$, {\color{blue}int}* \emph{xadj}, {\color{blue}int}* \emph{adjncy}, {\color{blue}bool} \emph{suppress\_output}, {\color{blue}int} \emph{seed}, {\color{blue} int} \emph{mode}, {\color{blue}int}** \emph{ordering}); \\

\subsubsection*{Parameters}
\begin{tabularx}{\textwidth}{lX}
$n$ & Number of vertices of the graph.\\
\emph{xadj} & This is the xadj array described in Section~\ref{ss:gds} which holds the pointers to the adjacency lists of the vertices. The array should have size $n+1$.\\
\emph{adjncy} & This is the adjacency array described in Section~\ref{ss:gds} which holds the adjacency lists of the vertices. The array should have size $2m$.\\
\emph{suppress\_output} & If this option is enabled then no output of the partitioning library is printed to stdout.\\
\emph{seed} & Seed to use for the random number generator.\\
\emph{mode} & One out of FAST, ECO, STRONG, FASTSOCIAL, ECOSOCIAL, STRONGSOCIAL. .\\
\emph{ordering} & This is an \emph{output} parameter. After the function call this array contains number of the node in the ordering.  \\
\end{tabularx}
\pagebreak
\subsubsection*{Process Mapping}
{\color{blue}void} process\_mapping({\color{blue}int}* $n$, {\color{blue}int}* \emph{vwgt}, {\color{blue}int}* \emph{xadj}, {\color{blue}int}* \emph{adjcwgt}, {\color{blue}int}* \emph{adjncy},\\  
\hspace*{2.925cm} {\color{blue}int}* \emph{hierarchy\_parameter},  {\color{blue}int}* \emph{distance\_parameter}, {\color{blue}int} \emph{hierarchy\_depth}, \\
\hspace*{2.925cm}  {\color{blue}double}* \emph{imbalance},  {\color{blue}bool} \emph{suppress\_output}, {\color{blue}int} \emph{seed}, {\color{blue} int} \emph{mode\_partitioning}, {\color{blue} int} \emph{mode\_mapping},\\
 \hspace*{3cm}{\color{blue}int}* \emph{edgecut}, {\color{blue}int}* \emph{qap}, {\color{blue}int}* \emph{part}); \\
        
\subsubsection*{Parameters}
\begin{tabularx}{\textwidth}{lX}
$n$ & Number of vertices of the graph.\\
\emph{vwgt} & This is the vertex weight array described in Section~\ref{ss:gds}. The array should have size $n$. If your graph does not have vertex weight you can use a null pointer as an argument.\\
\emph{xadj} & This is the xadj array described in Section~\ref{ss:gds} which holds the pointers to the adjacency lists of the vertices. The array should have size $n+1$.\\
\emph{adjcwgt} & This is the adjacency weight array described in Section~\ref{ss:gds} which holds the weights of the edges if they exists. The array should have size $2m$. If your graph does not have edge weights you can use a null pointer as an argument.\\
\emph{adjncy} & This is the adjacency array described in Section~\ref{ss:gds} which holds the adjacency lists of the vertices. The array should have size $2m$.\\
\emph{hierarchy\_parameter} & This is the array that describes the system hierarchy. For example, if you have 2 cores per node, 3 nodes per rack, and 4 racks, the array would have size 3 and the contents would be \{2,3,4\}. The array should have size as specified in \emph{hierarchy\_depth}.\\
\emph{distance\_parameter} & This is the array that describes the distances between the different components of the hierarchy. In the above example, the array should also have size 3 and look similar to \{1,10,100\}. The array should have size as specified in \emph{hierarchy\_depth}.\\
\emph{hierarchy\_depth} & This parameter described the depth of the hierarchy and hence the size of the previous two arrays specified.\\
\emph{imbalance} & This parameter controls the amount of imbalance that is allowed. For example, setting it to 0.03 specifies an imbalance of 3\% which means on unweighted graphs that each block has to fulfill the constraint $|V_i| \leq (1+0.03)|V|/k$.\\
\emph{suppress\_output} & If this option is enabled then no output of the partitioning library is printed to stdout.\\
\emph{seed} & Seed to use for the random number generator.\\
\emph{mode\_partitioning} & One out of FAST, ECO, STRONG, FASTSOCIAL, ECOSOCIAL, STRONGSOCIAL. Configuration names correpond to the default configuration names of the multilevel partitioner KaFFPa.\\
\emph{mode\_mapping} & One out of MAPMODE\_MUTLISECTION, MAPMODE\_BISECTION. We recommend MULTISECTION.\\
\emph{edgecut} & This is an \emph{output} parameter. It represents the edge cut of the computed partition.  \\
\emph{qap} & This is an \emph{output} parameter. It represents the value of the QAP objective function of the computed partition.  \\
\emph{part} & This is an \emph{output} parameter. It has to be an already allocated array of size $n$. After the function call this array contains the information of the blocks of the vertices, \ie the block of the $i$th vertex is given in \emph{part}[$i$].  \\
\end{tabularx}
\pagebreak
\subsubsection*{Using KaHIP within Java Projects}
KaHIP can be used within Java project via JNI. A small example on how to do this is provided in the folder \emph{misc/java\_jni\_wrapper/} of the project. To run the example perform the following commands in the main folder: \\

\noindent \emph{./compile\_withcmake.sh} \\
\emph{cd misc/java\_jni\_wrapper} \\
\emph{./makeWrapper.sh} \\
\emph{java KaHIPWrapper}
\vfill
\pagebreak
\bibliographystyle{plain}
\bibliography{phdthesiscs}
\end{document}